\documentclass{article}

\usepackage{epsfig}
\usepackage{amsmath}
\usepackage{amsfonts}
\usepackage{graphicx}

 \usepackage{amssymb}

\date{\today}

\newcommand{\insertplot}[5]{\begin{figure}
 \hfill\hbox to 0.05in{\vbox to #5in{\vfill
 \inputplot{#1}{#4}{#5}}\hfill}
 \hfill\vspace{-.1in}
 \caption{#2}\label{#3}
 \end{figure}}
 \newcommand{\inputplot}[3]{
 \special{ps: plotfile #1}
}
\newcounter{fig}

\newcommand{\vep}{\varepsilon}
\newcommand{\nn}{\nonumber}

\newcommand{\la}{\lambda}

\newcommand{\Ga}{\Gamma}

\newcommand{\si}{\sigma}
\newcommand{\Si}{\Sigma}

\newcommand{\ee}{\end{equation}}
\newcommand{\eea}{\end{eqnarray}}
\newcommand{\be}{\begin{equation}}
\newcommand{\bea}{\begin{eqnarray}}

\def\theequation{\arabic{equation}}

\newcommand{\re}[1]{(\ref{#1})}
\newcommand{\R}{{\rm I \hspace{-0.52ex} R}}

\begin{document}

\title{
%Dyon-like solutions
Non-Abelian clouds around Reissner-Nordstr\"om black holes:
\\
the existence line
}
 \vspace{1.5truecm}
\author{
{\large Eugen Radu}$^{\dagger}$,
   {\large D. H. Tchrakian}$^{\star \diamond }$
	and {\large Yisong Yang}$^{\ddagger}$
	%\, \triangle}$
	%
\vspace*{0.2cm}
\\
$^{\dagger}${\small Departamento de Fisica da Universidade de Aveiro and CIDMA,
 Campus de Santiago, 3810-183 Aveiro, Portugal}
   \\
$^{\star}${\small
School of Theoretical Physics -- DIAS, 10 Burlington
Road, Dublin 4, Ireland} \\
$^{\diamond}${\small  Department of Computer Science,
National University of Ireland Maynooth,
Maynooth,
Ireland}
\\
$^{\ddagger}${\small  Department of Mathematics, Tandon School, New York University, Brooklyn, New York 11201, USA} 
}

\maketitle

\begin{abstract}

A known feature of electrically charged Reissner-Nordstr\"om-anti-de Sitter 
planar black holes is that they can become unstable
when considered as solutions of Einstein--Yang-Mills theory.
The mechanism for this is that the linearized Yang-Mills
equations in the background of the Reissner-Nordstr\"om (RN) black holes
possess a normalizable zero mode, resulting in non-Abelian (nA) magnetic clouds near the horizon.
In this work we show that the same pattern may occur also for
asymptotically flat RN black holes.
Different from the anti-de Sitter  case, in the Minkowskian background
the prerequisite for the existence of the nA clouds is $i)$
a large enough gauge group  and
$ii)$ the presence of some
extra interaction terms in the matter Lagrangian.
To illustrate this mechanism
we present two specific examples, 
one in four and the other in five dimensional asymptotically flat spacetime.
In the first case, we augment the usual $SU(3)$ Yang-Mills
Lagrangian with higher order (quartic) curvature term,
while for the second one we add the Chern-Simons density to the $SO(6)$ Yang-Mills system.
In both cases, an Abelian
gauge symmetry is spontaneously broken near a RN black
hole horizon with the appearance of a condensate of nA gauge fields.
In addition to these two examples, we review the corresponding picture for
anti-de Sitter black holes.
All these solutions are studied both analytically and numerically,
existence proofs being provided for nA clouds in the background of RN black holes.
The proofs use  shooting techniques which are suggested by and in turn offer insights
for our numerical methods.
They indicate that, for a black hole of given mass, appropriate electric charge values 
are required to ensure the existence
of solutions interpolating desired boundary behavior at the horizons and spatial infinity.

\end{abstract}

%%%%%%%%%%%%%%%%%%%%%%%%%%%%%%%%%%%%%%%%%%%%%%%%%%%%%%%%%%%%%%%%%%%%%%%%%%%%%%%
\section{Introduction}
%%%%%%%%%%%%%%%%%%%%%%%%%%%%%%%%%%%%%%%%%%%%%%%%%%%%%%%%%%%%%%%%%%%%%%%%%%%%%%%
The branching off of a family of solutions of a model into a
new family of solutions at the onset of an instability is a
recurrent situation in physics.
Starting with vacuum Einstein theory of gravity,
an earlier example is the Gregory-Laflamme instability \cite{Gregory:1993vy}
of black strings, which branch off
to a family of nonuniform string solutions at the onset of the
instability \cite{Gubser:2001ac}.
A more recent example is the bumpy black holes (BHs),
which branch off the  family of Myers-Perry vacuum BHs
in $d\geq 6$ spacetime dimensions
\cite{Dias:2009iu},
\cite{Dias:2014cia}.

The same pattern occurs, however, for some
field theory models in a flat spacetime background,
the best known example being perhaps the bisphalerons in
the electroweak sector of the Standard Model of particle physics
\cite{Kunz:1988sx},
\cite{Yaffe:1989ms}.
As expected, even more complicated solutions
are found when considering gravitating matter fields,
with new features introduced by the possible existence of an event horizon.
A recent example in this direction is
found for a complex massive scalar field in the background
of a Kerr BH.
As discussed in
\cite{Hod:2012px},
\cite{Herdeiro:2014goa},
\cite{Benone:2014ssa},
the Klein-Gordon equation possesses bound state solutions ($i.e.$ scalar clouds) around Kerr BHs 
(see also \cite{Li:2015bfa}
for a closely related system).
This leads to the existence of a family
of hairy BHs which branch off from the Kerr metric at the threshold of the superradiant instability
\cite{Herdeiro:2014goa}
(the existence of these scalar clouds at the nonlinear level has been proven recently in
\cite{Chodosh:2015oma}).

 \medskip
An important example
 which has recently received a considerable
amount of interest,
concerns the instability of the (electrically charged) Reissner-Nordstr\"om
(RN) BH with a negative cosmological constant
when considered as a solution of the gravitating $U(1)$-gauged complex scalar field theory
\cite{Gubser:2008px},
or of the Einstein--Yang-Mills (EYM) theory
\cite{Gubser:2008zu}.
In both cases, the branching towards a set of  solutions in
the more general theory occurs for a particular set of Reissner-Nordstr\"om-anti de Sitter (RNAdS)
BHs which form a line in the parameter space.
The hairy BHs are the nonlinear realization of those marginally stable modes.
This has led to the discovery of a
remarkable connection between condensed matter and gravitational physics,
 the hairy black hole solutions in the aformentioned theories
being interpreted as {\it "holographic superconductors"},
via the gauge/gravity duality \cite{Horowitz:2010gk}.

 \medskip
The main purpose of the present work is to
show that the mechanism resulting in the spontaneous breaking of an Abelian
gauge symmetry near a black
hole horizon, with the appearance of a condensate of magnetic non-Abelian (nA) gauge field,
also occurs  for gravitating nA fields  in the absence of a cosmological constant~\footnote{Note that the (asymptotically flat,
magnetically charged) RN BH
is known to be unstable when viewed as a solution of the Einstein--Yang-Mills--Higgs theory \cite{Lee:1991qs}.}.
Following the terminology for scalar fields,
\cite{Hod:2012px},
\cite{Herdeiro:2014goa},
\cite{Benone:2014ssa},
 these configurations with infinitesimally small magnetic fields are dubbed here
{\it non-Abelian clouds}.
Moreover, the corresponding bifurcating Reissner-Nordstr\"om (RN) BHs will correspond to an {\it existence line}
in the parameter space of solutions. 

There is, however, a price to be paid in this case.
Firstly, the gauge group must be larger
than $SO(d-1)$ (with $d$ the spacetime dimension), and second, a suitable YM additional interaction term,
 with its corresponding dimensional constant,
must be employed.
There are two candidates for this, the nonlinear densities constructed from the nA fields and connections, namely
$i)$ as
many as allowed higher-order curvature terms in the given dimension, and 
$ii)$ the allowed
Chern-Simons (CS) density in that (odd) dimension.

For economy of presentation, we have chosen to demonstrate this effect for two typical examples,
in four and five spacetime dimensions
respectively. In the $d=4$ case, we do not have the option of employing a CS term, and the only higher order
curvature term available there is the quartic kinetic YM term $F^4=\mbox{Tr}\{F_{\mu\nu\rho\si}F^{\mu\nu\rho\si}  \}$.
As for gauge group, we have chosen $SU(3)$, which is the
smallest group larger than $SO(4)$ or $SU(2)$.
In five dimensions however, where we have available the CS density,
we have eschewed the use of the $F^4$ form for simplicity,
and have chosen to employ the CS term, for diversity. In that case, we have
chosen to work with the gauge group $SO(6)$ for convenience.

%\medskip
 
With these two examples, we aim to illustrate this mechanism for
generic cases.
Let us also mention that
our study uses 
a combination of analytical and numerical methods,
which is enough for most purposes.
In particular, we have given
analytic proofs of existence in addition to numerical constructions. 

The methods of proofs are based on shooting arguments 
utilizing the black hole electric charge as shooting parameter
which gives rise to the boundary slope of the gauge field profile function at the horizon. 
In order to obtain correct values of the
charge, two steps of shooting processes are conducted. 
These methods are hinted by and useful to the numerical approaches employed in this study.

\medskip
Our work is organized as follows.
Before considering the case of main interest of solutions in a Minkowski
spacetime background,
 we start in Section 2 by reviewing the mechanism unveiled in \cite{Gubser:2008zu}
for the  
 occurrence of nA clouds in the
 gravitating $SU(2)$ YM system with a negative cosmological constant. 
In Sections 3 and 4,
we discuss two specific examples for the instability of asymptotically flat
RN BH with nA fields, in four and five dimensions.
In the first case, one supplements the usual Yang-Mills
Lagrangian with higher order curvature terms of the gauge field, while for the latter
one considers the Yang-Mills--Chern-Simons theory.
The numerical results there are
underpinned by rigorous existence proofs for the solutions of the corresponding
linearized (generalized) Yang-Mills equations.
We end with Section 5, where the results are summarised and discussed.
The Appendix contains essentially the equations for the full non-linear systems
discussed in Sections 2-4, put, however, in a more general context.

%%%%%%%%%%%%%%%%%%%%%%%%%%%%%%%%%%%%%%%%%%%%%%%%%%%%%%%%%%%%%%%%%%%%%%%%%%%%%%%
\section{Non-Abelian clouds around $d=4$ Reissner-Nordstr\"om--Anti-de Sitter black holes}
%%%%%%%%%%%%%%%%%%%%%%%%%%%%%%%%%%%%%%%%%%%%%%%%%%%%%%%%%%%%%%%%%%%%%%%%%%%%%%%

%%%%%%%%%%%%%%%%%%%%%%%%%%%%%%%%%%%%%%%%%%%%%%%%%%%%%%%%%%%%%%%%%%%%%%%%%%%%%%%
 \subsection{The setting}
%%%%%%%%%%%%%%%%%%%%%%%%%%%%%%%%%%%%%%%%%%%%%%%%%%%%%%%%%%%%%%%%%%%%%%%%%%%%%%%
We consider the usual EYM action supplemented with a cosmological term $\Lambda=-3/L^2$
\bea
\label{action-AdS}
S=\int d^4 x \sqrt{-g}
\left [\frac{1}{4}  (R+\frac{6}{L^2})
-\frac{1}{2}\mbox{Tr}  \{F^2 \}
\right] .
\eea
where the field strength tensor $F_{\mu \nu}=\frac{1}{2} \tau^a F_{\mu\nu}^{a}$
 is
%\begin{equation}
$
F_{\mu \nu} =
\partial_\mu A_\nu -\partial_\nu A_\mu + i \left[A_\mu , A_\nu \right],
$
%\ , \label{fmn}
%\end{equation}
%F_{\mu \nu}^a =
%\partial_\mu A_\nu^a -\partial_\nu A_\mu^a + i\epsilon^{abc}A_\mu^b A_\nu^c
and the gauge field
$A_{\mu} = \frac{1}{2} \tau^a A_{\mu}^{a},$ with $\tau^a$ an SU(2) basis
written in terms of Pauli matrices.

The  RNAdS BH
with a planar horizon is
a solution of this model and has a line element
\begin{eqnarray}
\label{RNAdS}
ds^{2}=\frac{dr^{2}}{N(r)}+r^2 (d\theta^2+\theta^2 d \varphi^2)- N(r) dt^{2},~~
{\rm with}~~N(r)=-\frac{2M}{r}+\frac{Q^2}{r^2}+\frac{r^2}{L^2}
\end{eqnarray}
(with $M$, $Q$ two constants fixing the mass and electric charge of the BH)
and a purely electric gauge field,
\begin{eqnarray}
\label{RNAdS-A}
A=V(r)\frac{1}{2}\tau_3 dt,~~~{\rm with}~~~V(r)= \left( \frac{Q}{r_H}-\frac{Q}{r} \right)dt,
\end{eqnarray}
where $r_H$ is the largest root of the equation $N(r_H)=0$.
In the above relations, $r,t$ are the radial and time coordinates, respectively,
while $\theta$ and $\varphi$ are coordinates on the two-plane, with
$0\leq \theta <\infty$,
$0\leq \varphi <2\pi$.

The genuine nA solutions are found when exciting the magnetic
degrees of freedom of the SU(2) potential, with 
\cite{VanderBij:2001ia},
\cite{Mann:2006jc}
\begin{eqnarray}
\label{RNAdS-A2}
A=V(r)\frac{1}{2}\tau_3 dt+\frac{1}{2} \Big\{
w(r) \tau_1  d \theta
+\big( \tau_3
+ w(r) \tau_2 \theta \big)  d \varphi \Big\},
\end{eqnarray}
and would describe dyonic nA BHs
(the corresponding equations of the model are given in Appendix A1).
Here we are interested in the case when $w(r)$ is an infinitesimally small function
\begin{eqnarray}
\label{1}
w(r)=\epsilon W(r),
\end{eqnarray}
such that the backreaction induced by the magnetic field on the
spacetime geometry (\ref{RNAdS}) can be neglected and
 the
  RNAdS BH remains a solution of the model.
	Then the function
  $W(r)$ solves the linearized YM equation
\begin{eqnarray}
\label{eq-AdS}
W''+\frac{N'}{N}W'+\frac{\left( \frac{Q}{r_H}-\frac{Q}{r} \right)^2}{N^2}W=0,
\end{eqnarray}
 where a prime denotes a derivative $w.r.t.$ the coordinate $r$.
%Finally, let us mention that the existence of an is a feature of AdS black holes with a planar horizon topology.

%%%%%%%%%%%%%%%%%%%%%%%%%%%%%%%%%%%%%%%%%%%%%%%%%%%%%%%%%%%%%%
\subsection{The parameter dependence of solutions}
%%%%%%%%%%%%%%%%%%%%%%%%%%%%%%%%%%%%%%%%%%%%%%%%%%%%%%%%%%%%%%
The above equation does not seem to possess a closed form solution.
One can construct, however,  approximate solutions valid for $r\to r_H$ and $r\to \infty$,
respectively.
In the
vicinity of the event horizon, the regular solution has
\begin{eqnarray}
\label{3}
W(r)=b+ \frac{L^4Q^2r_H^2 w_0}{4 (L^2Q^2-3r_H^4)^2} (r-r_H)^2+\dots,
\end{eqnarray}
 (with $b\neq 0$ arbitrary),
the corresponding expression for large-$r$ being
\begin{eqnarray}
\label{4}
W(r)=\frac{J}{r}-\frac{L^4Q^2 J}{6r_H^2 } \frac{1}{ r^3} +\dots\ .
\end{eqnarray}

In what follows we investigate the existence of a smooth solution
connecting these asymptotics.
Since (\ref{eq-AdS}) is linear and homogeneous,
we may set $b=1$ without loss of generality and look for a solution satisfying $W(\infty)=0$.
For this purpose, we rewrite $N(r)$ as
\be
\nonumber
N(r)=\frac1{r^2}\left(\frac1{L^2}r^4-2Mr+Q^2\right)\equiv\frac1{r^2} f(r),\quad r>0,
\ee
and locate $r_H$ first. It is seen that the only root of $f'(r)$ is
%\be
$
r_0=2^{-\frac13}M^{\frac13}L^{\frac23}.
$
%\ee
In order that $N(r)$ has a positive root it is necessary and sufficient to require $f(r_0)\leq0$ which leads to the condition
\be
Q^2\geq 2^{\frac43} M^{\frac43}L^{\frac23}\left(1-2^{-\frac83}\right)\equiv Q^2_0,\quad Q_0>0.
\ee
If $Q=Q_0$, $N(r)$ will have a double root at $r=r_0$; if $Q>Q_0$, $N(r)$ will have two positive simple roots which both depend on $Q$ and may be denoted as
$r_1(Q)$ and $r_2(Q)$ with $r_1(Q)<r_2(Q)$. Thus in our notation $r_H=r_2(Q)$. Since $f(r)>-2Mr +Q^2$, we have
\be
\nonumber
\frac{Q^2}{2M}<r_1(Q)<r_2(Q)=r_H(Q).
\ee

Suggested by the asymptotic expression (\ref{3}), we are to obtain a positive solution of (\ref{eq-AdS}) subject to the boundary condition
\be
W(r_H)=1,\quad W(\infty)=0,
\ee
following from (\ref{3}), (\ref{4}).
With this, we may begin by integrating (\ref{eq-AdS}) to represent the unique local solution with the initial condition $W(r_H)=1,W'(r_H)=0$ as
\be\label{x15}
W'(r)=-\frac1{N(r)}\int_{r_H}^r\frac{Q^2}{N(\rho) r_H^2\rho^2}(\rho-r_H)^2 W(\rho)\,d\rho,\quad r>r_H.
\ee
Integrating (\ref{x15}) we formally arrive at
\be\label{xx15}
W(r)=1-\int_{r_H}^r \frac1{N(s)}\int_{r_H}^s\frac{Q^2}{N(\rho) r_H^2\rho^2}(\rho-r_H)^2 W(\rho)\,d\rho\,ds,\quad r>r_H.
\ee
The parameter dependence of the solution is demonstrated in (\ref{x15}) and (\ref{xx15}) which allows us to obtain desired solution profiles
by parameter adjustment.

%%%%%%%%%%%%%%%%%%%%%%%%%%%%%%%%%%%%%%%%%%%%%%%%%%%%%%%%%%%%%%%%%%%%%%%%%%%%%%%
 \subsection{The numerical results}
%%%%%%%%%%%%%%%%%%%%%%%%%%%%%%%%%%%%%%%%%%%%%%%%%%%%%%%%%%%%%%%%%%%%%%%%%%%%%%%

In practice,
the solution interpolating between the  asymptotics
(\ref{3}),
(\ref{4})
is found numerically, a typical profile being shown in Figure 1.
Note that we restrict here to configurations with $W(r)$ everywhere positive; 
however, there are also excited solutions in which $W(r)$ has nodes \cite{Gubser:2008zu}.

The results of the numerical integration   are shown in Figure 2,
where we exhibit the existence line in the $(M,Q)$-parameter space of RNAdS solutions
(note that there we set $L=1$).
One can see that, given a value of the electric charge $Q$, the (nodeless) solution of (\ref{eq-AdS})
is found for a critical RNAdS BH only.

As discussed in \cite{Gubser:2008zu},
when taking into account the backreaction at the nonlinear level,
 this results  in the occurrence of a branch of EYM hairy black hole solutions
which bifurcate from the RNAdS configurations precisely at the existence line.
Moreover, the solutions with a nonzero magnetic potential are thermodynamically favoured
over the magnetically neutral ones
($i.e.$ they maximize the entropy for given $M,Q$).

%%%%%%%%%%%%%%%%%%%%%%%%%%%%%%%%%%%%%%%%%%%%%%%%%%%%%%%%%%%
 \setlength{\unitlength}{1cm}
\begin{picture}(8,6)
\put(3,0.0){\epsfig{file=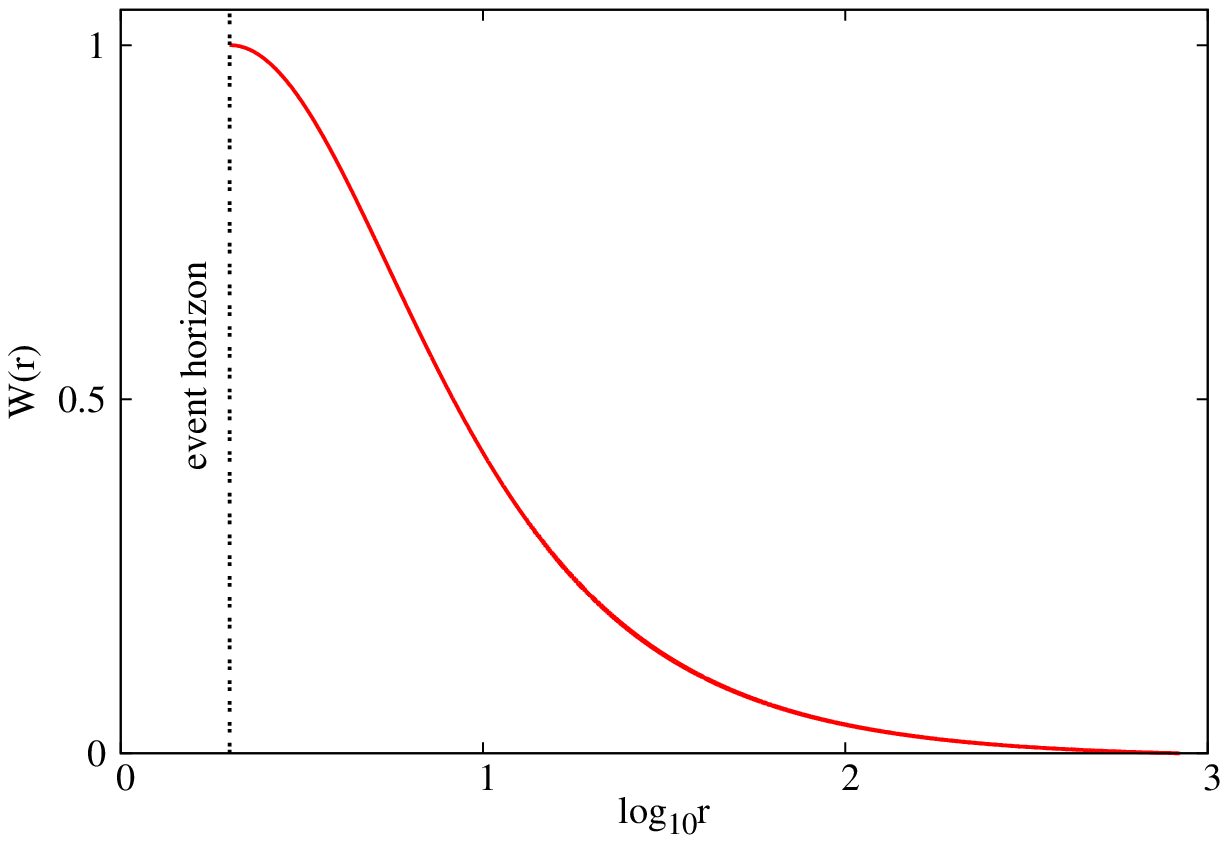,width=8cm}}
\end{picture}
\\
\\
{\small {\bf Figure 1.}
The profile of a typical solution of the equation (\ref{eq-AdS}).
The parameters of the corresponding RNAdS background are $M=0.0453$, $Q=0.1465$ and $L=10$.
 }
 %\vspace{0.5cm}

%%%%%%%%%%%%%%%%%%%%%%%%%%%%%%%%%%%%%%%%%%%%%%%%%%%%%%%%%%%
 \setlength{\unitlength}{1cm}
\begin{picture}(8,6)
\put(3,0.0){\epsfig{file=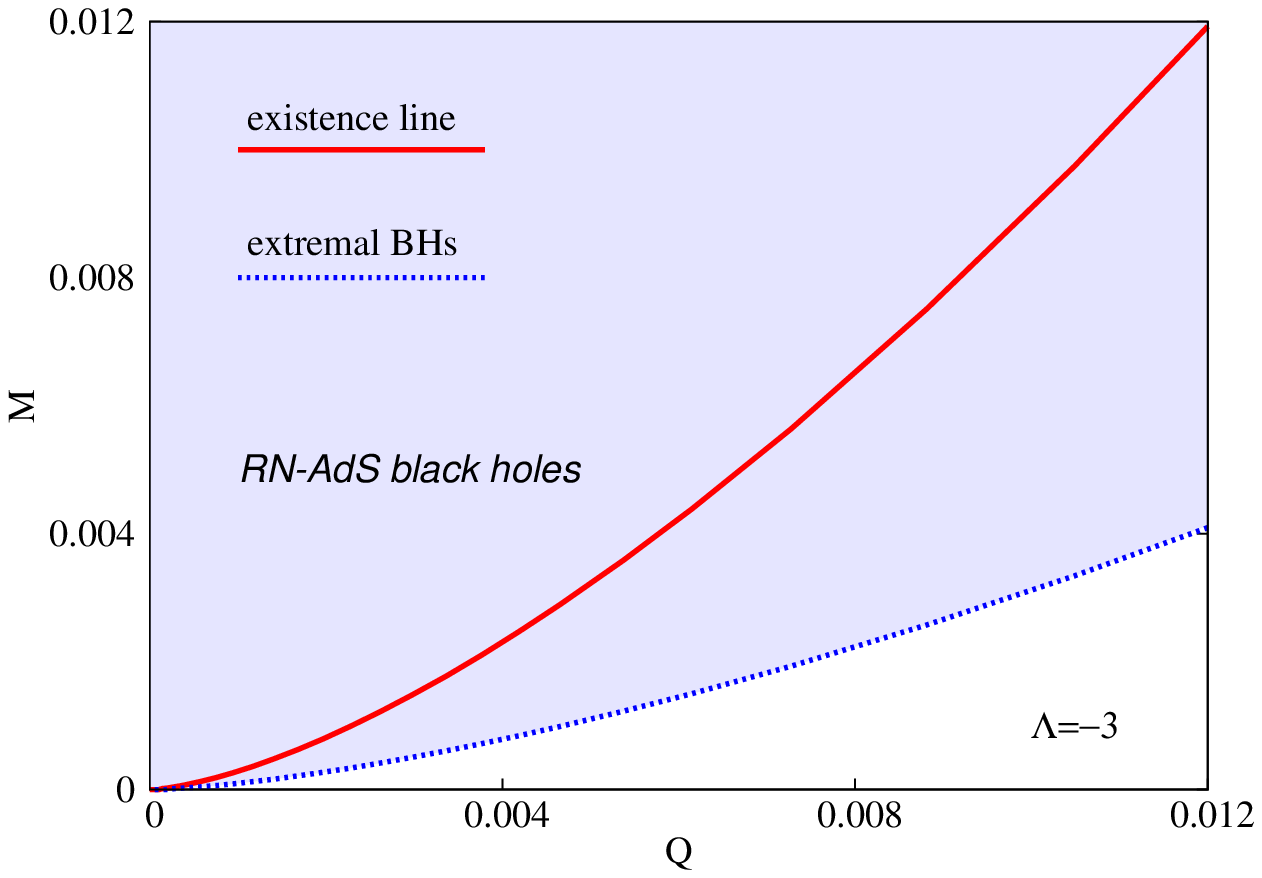,width=8cm}}
\end{picture}
\\
\\
{\small {\bf Figure 2.}
Mass $M$ $vs.$ charge $Q$ for RNAdS black holes in $d=4$
dimensions. The dotted blue
 curve corresponds to extremal BHs
(RN BHs exist above it (shaded region).
The non-Abelian clouds exists along the red line.
 }
%\vspace{0.5cm}
%%%%%%%%%%%%%%%%%%%%%%%%%%%%%%%%%%%%%%%%%%%%%%%%%%%%%%%%%%%

%%%%%%%%%%%%%%%%%%%%%%%%%%%%%%%%%%%%%%%%%%%%%%%%%%%%%%%%%%%%%%%%%%%%%%%%%%%%%%%
\section{Non-Abelian clouds around $d=4$ Reissner-Nordstr\"om black holes
with Minkowski spacetime asymptotics}
%%%%%%%%%%%%%%%%%%%%%%%%%%%%%%%%%%%%%%%%%%%%%%%%%%%%%%%%%%%%%%%%%%%%%%%%%%%%%%%

%%%%%%%%%%%%%%%%%%%%%%%%%%%%%%%%%%%%%%%%%%%%%%%%%%%%%%%%%%%%%%%%%%%%%%%%%%%%%%%
 \subsection{The setting}
%%%%%%%%%%%%%%%%%%%%%%%%%%%%%%%%%%%%%%%%%%%%%%%%%%%%%%%%%%%%%%%%%%%%%%%%%%%%%%%

It is worth inquiring to which extent the features unveiled above are specific
to AdS spacetime, and whether they can be recovered (at least to some extent)
 also by asymptotically flat configurations.
The first observation is that in this case the BHs
necessarily possess a spherical horizon topology \cite{HE}, \cite{Friedman:1993ty}.
Second, for $\Lambda=0$, the static configurations with
a gauge group $SU(2)$ and a non-trivial magnetic potential necessarily have
$A_t \equiv 0$ as found by
a number of classic no go ``{\it baldness}'' theorems \cite{bald}.
This immediately excludes the existence of $SU(2)$ clouds around
electrically charged Reissner-Nordstr\"om black holes.

One may hope that the situation changes when considering instead a larger gauge group.
The minimal gauge group for which the
superposition of a Coulomb field
and a non-Abelian hair is not forbidden by the ``{\it baldness}'' theorems is $SU(3)$.
In this case, a spherically symmetric Ansatz for the gauge field reads\footnote{Note that (\ref{YMansatz}) 
corresponds in fact to an $SU(2)\times U(1)$ Ansatz,
the magnetic and electric potential interacting only via the spacetime geometry.} \cite{Galtsov:1991au}
\begin{eqnarray}
\label{YMansatz}
A=
  w(r) T_1  d\theta
 +\left(  w(r) T_2  \sin \theta +\cos \theta T_3 \right) d\varphi
 + V(r) T_8  dt ,
\end{eqnarray}
where $T_i$ are the standard generators of the $SU(3)$ Lie algebra.
The electrically charged RN BH remains a solution of the EYM-SU(3) model, with a line element
\bea
\label{RN2}
ds^{2}=\frac{dr^{2}}{N(r)}+r^2 (d\theta^2+\sin^2 \theta^2 d \varphi^2)- N(r) dt^{2},~~
{\rm with}~~N(r)=1-\frac{2M}{r}+\frac{Q^2}{r^2} ,
\eea
(with $M$ and $Q$ the mass and electric charge of BHs)
and an embedded Abelian connection
\bea
\label{n2}
w(r)=\pm 1,~~
V(r)= \left( \frac{Q}{r_H}-\frac{Q}{r} \right)dt,~
\eea
$r_H=M+\sqrt{M^2-Q^2}>0$ is an input parameter -- the outer event horizon (with $N(r_H)=0$).

Apart from that, there are also genuinely nA solutions which possess
a nontrivial magnetic potential $w(r)$.
As discussed in \cite{Galtsov:1991au},
these configurations can be thought of as nonlinear superpositions
of the (electric) RN and the (purely magnetic) SU(2) black holes \cite{89}.
In particular, they do not emerge as
perturbations of the electrically charged BHs,
the overall picture being very different from that in the AdS case.

These results are found for a usual YM action which contains the usual quadratic term $F_{\mu \nu}F^{\mu\nu}$  only.
However, as discussed in \cite{Radu:2011ip},
the situation changes  when
the nA action is augmented with higher order curvature terms of the gauge field.
In the simplest case the Lagrangian for the SU(3) fields reads
\begin{eqnarray}
\label{L}
{\cal L}= -\frac{1}{2 }{\rm Tr}\left \{ F_{\mu\nu}F^{\mu\nu} \right\}+{\cal L}_s,~~{\rm with}~~~
{\cal L}_s=\frac{3\tau}{2 }
 {\rm Tr}\left\{ (F_{\mu \nu} {}\tilde F^{\mu \nu})^2 \right \},
\end{eqnarray}
where a tilde denotes the Hodge dual and $\tau$
is an input parameter of the theory.
One can see that  $\mathcal{L}_s$ features only the second power of any {\it ``velocity field''} and
is a causal density just
like the Gauss-Bonnet term in gravity \cite{Zwiebach:1985uq} or the
Skyrme~\cite{Skyrme:1962vh}
term of the $O(4)$ sigma model. Also, $\mathcal{L}_s$
can be viewed as the
second member of the YM hierarchy \cite{Tchrakian:1984gq},
providing a natural generalization of the usual YM model.
Such terms were extensively considered in the literature for Abelian solutions;
however, the nA case is considerably less studied.

The equations of motion for a spherically symmetric system
are displayed in Appendix A2.
An important observation is that the (electrically charged) RN BH
is still a solution of this model with the extra ${\cal L}_s$-term.
We are interested in the configurations with
 $w^2(r)$ infinitesimally close to $1$,
such that the magnetic part of the
nA field is too small to back-react significantly upon the geometry,
\begin{eqnarray}
\label{p}
w(r)=-1+\epsilon W(r),
\end{eqnarray}
in which case the geometry is still 
described by the metric functions in (\ref{RN2}),
sourced by a purely electric YM field.

Then the linearized YM equations in A2 imply that $W(r)$
solves the equation
\begin{eqnarray}
\label{4.1}
(NW')'=\frac{2W}{r^2}(1-\frac{2Q^2\tau}{r^4}).
\end{eqnarray}
  %

%%%%%%%%%%%%%%%%%%%%%%%%%%%%%%%%%%%%%%%%%%%%%%%%%%%%%%%%%%%%%%
\subsection{The existence of  solutions}
%%%%%%%%%%%%%%%%%%%%%%%%%%%%%%%%%%%%%%%%%%%%%%%%%%%%%%%%%%%%%%

We are interested in
the smooth solutions of (\ref{4.1}) 
with the following form  as $r\to r_H$
\begin{eqnarray}
\label{n0}
W(r)= b+w_1(r-r_H)+O(r-r_H)^2,
\end{eqnarray}
where
\begin{eqnarray}
\nonumber
w_1=\frac{2b(r_H^4-2Q^2\tau)}{r_H^3(r_H^2-Q^2)},
\end{eqnarray}
with $b\neq 0$ an arbitrary parameter.
Also, since the eq. (\ref{4.1}) is linear, we set $b=1$
without any loss of generality.
The corresponding expression for large $r$ reads
\begin{eqnarray}
\label{n1}
W(r)=\frac{J}{r}+\frac{3J(Q^2+r_H^2)}{4r_H}\frac{1}{r^2}+O(1/r^3),
\end{eqnarray}
with $J$ a constant.

The existence of a smooth solution interpolating between these asymptotics
can be shown as follows.
First, we fix the length scale of the problem by taking $r_H=1$, such that
\be
 N(r)=\left(1-\frac1r\right)\left(1-\frac{Q^2}r\right),\quad 0<Q<1,\quad r\geq1.
\ee
From (\ref{n0}), (\ref{n1}),
the solutions of (\ref{4.1}) are
subject to the boundary condition
\be\label{4.3}
W(1)=1,\quad W(\infty)=0.
\ee
It is easy to see that (\ref{4.1}) has a unique local solution satisfying the initial condition
\be\label{4.4}
W(1)=1,\quad W'(1)=a,\quad a\in\R.
\ee
For convenience, we denote such a solution as $W(r;a)$ and define the set
\be
\nonumber
S^+=\{a\in\R\,|\, W'(r;a)=W_r(r;a)>0\mbox{ for some }r>1\}.
\ee
It is clear that $S^+$ is open.
It is trivial that $(0,\infty)\subset S^+$.

We recast (\ref{4.1}) into
\be\label{4.6}
W'(r;a)=\frac1{(r-1)}\frac{r^2}{(r-Q^2)}\int_1^r\frac2{\rho^2}\left(1-\frac{2Q^2\tau}{\rho^4}\right)W(\rho;a)\,d\rho,\quad r>1.
\ee
Letting $r\to1$ in (\ref{4.6}) and applying L'Hospital's rule, we find
\be \label{4.7}
a=\lim_{r\to1}W'(r;a)=\frac{2(1-2Q^2\tau)}{(1-Q^2)}.
\ee
So we may adjust $Q$ in the interval $(0,1)$ to make $a>0$ or $a\in S^+$.

In order to have a solution with $a<0$, we need to request
\be\label{4.8}
2Q^2\tau>1,
\ee
which will be observed in the sequel.

We next consider the interval
\be
I_0=[1,r_0]\equiv \left[1,(2Q^2\tau)^{\frac14}\right].
\ee
From (\ref{4.6}) we see that $W'<0$ provided that $r\in I_0$ and $W$ stays non-negative for $r\in I_0$. With this condition, we have
\bea
W'(r;a)&>&\frac{r^2}{(r-1)(r-Q^2)}\int_1^r\frac2{\rho^2}\left(1-\frac{r_0^4}{\rho^4}\right)\,d\rho\nn\\
&>&\frac{2r_0^2}{5(r_0-Q^2)}(r^4-r_0^4),\quad r\in I_0=[1,r_0].
\eea
Integrating the above, we get the lower bound
\bea
W(r;a)&>&1+\frac{2r_0^2}{5(r_0-Q^2)}\left(\frac15[r^5-1]-r_0^4[r-1]\right)\nn\\
&\geq&1+\frac{2r_0^2}{5(r_0-Q^2)}\left(\frac15[r_0^5-1]-r_0^4[r_0-1]\right),\quad r\in I_0,
\eea
which may be made positive by adjusting $Q$ in the interval (\ref{4.8}) suitably.

Similarly, we can find $a$ suitably such that $W(r;a)<0$ for some $r\in (1,r_0)$. We set
\bea
\nonumber
A^+&=&\{a |\, W(r;a)>0\mbox{ for all }r\in I_0\},
\\
\nonumber
A^-&=&\{a|\, W(r;a)<0\mbox{ for some } r\in I_0\}.
\eea
Then $A^+$ and $A^-$ are open sets which may be restricted to an open interval, say $I$, for $Q$ satisfying the condition (\ref{4.8}).
The connectedness of $I$ implies $A^0=I\setminus(A^+\cup A^-)\neq\emptyset$.

Take $a\in A^0$. Then $W(r;a)\geq0$ for all $r\in I_0$ and $W(r_1;a)=0$ for some $r_1\in I_0$. It is clear that $r_1=r_0$ otherwise
$r_1$ would be an interior minimum of $W$ over $I_0$ resulting in $W'(r_1;a)=0$, thus $W\equiv0$
by the uniqueness theorem of the initial value problem of ordinary
differential equations, which contradicts the initial condition $W(1;a)=1$. Thus we have
\be\label{4.14}
W(r;a)>0\quad\mbox{for }r\in [1,r_0);\quad W(r_0;a)=0.
\ee
Inserting (\ref{4.14}) into (\ref{4.6}), we obtain
\be\label{4.15}
W'(r;a)<0, \quad r\in I_0.
\ee

We claim $W'(r;a)<0$ for all $r>1$. In fact, for $r$ near and above $r_0$, we have from (\ref{4.6})
\be\label{4.16}
W'(r;a)=\frac1{(r-1)}\frac{r^2}{(r-Q^2)}\left(\int_1^{r_0}+\int_{r_0}^r\right)
\left(\frac2{\rho^2}\left[1-\frac{2Q^2\tau}{\rho^4}\right]W(\rho;a)\right)\,d\rho,
\ee
which will continue to stay negative for all $r>r_0$ since $W$ becomes negative beyond $r_0$.

Therefore, we have shown that the set
\be
\nonumber
S^-=\{a\in\R\,|\, W'(r;a)<0 \mbox{ for all } r>1\mbox{ and } W(r;a)<0\mbox{ for some }r>1\}
\ee
is not empty. This set is clearly open.

Using the connectedness of $\R$, we see that $S^0=\R\setminus(S^+\cup S^-)\neq\emptyset$. For $a\in S^0$, we have
\be\label{4.18}
W'(r;a)\leq 0,\quad W(r;a)\geq0,\quad r>1.
\ee
Applying the uniqueness theorem of the initial value problem of ordinary
differential equations again, we get $W>0$.

We claim  $W'<0$ for all $r>1$. In fact, we have $W'(r;a)<0$ for $r\in I_0$ since $W(r;a)>0$ for $r\in I_0$.
Assume otherwise that $W'$ has at least one zero and set
\be
\nonumber
r_1=\inf\{ W'(r;a)=0\,|\, r>r_0\}.
\ee
Then $r_1>r_0$ and
\be
\nonumber
\int_1^{r_1} \left(\frac2{\rho^2}\left[1-\frac{2Q^2\tau}{\rho^4}\right]W(\rho;a)\right)\,d\rho=W'(r_1;a)\frac{(r_1-1)(r_1-Q^2)}{r_1^2}=0.
\ee
Using this result in the decomposition (\ref{4.16}) with setting $r_0=r_1$ and applying $W>0$, we have
 \be\label{4.21}
W'(r;a)=\frac1{(r-1)}\frac{r^2}{(r-Q^2)}\int_{r_1}^r\frac2{\rho^2}\left(1-\frac{r_0}{\rho^4}\right)W(\rho;a)\,d\rho>0,
\ee
which is false. Thus we have strengthened (\ref{4.18}) into
\be\label{4.22}
W'(r;a)< 0,\quad W(r;a)>0,\quad \forall r>1.
\ee
With the properties of $W$ stated in (\ref{4.22}), we now establish the limit
\be\label{4.23}
\lim_{r\to\infty} W(r;a)\equiv W_\infty=0.
\ee
 
In fact, if $W_\infty\neq0$, then
\be\label{4.24}
W(r;a)>W_\infty>0,\quad \forall r>1.
\ee

From (\ref{4.6}) we see that
\be\label{4.25}
W_1\equiv \lim_{r\to\infty} W'(r;a)
\ee
exists. Since (\ref{4.23}) exists, we see that $W_1=0$ in (\ref{4.25}).  Hence, by  L'Hospital's rule and (\ref{4.1}), we have
\be
\lim_{r\to\infty}\frac{NW'(r;a)}{\left(-\frac1r\right)}=\lim_{r\to\infty} r^2 (NW'(r;a))'=2W_\infty.
\ee
 Since $W_\infty>0$, we can find some $r_2>0$ sufficiently large so that
 \be
\nonumber
 \frac{NW'}{\left(-\frac1r\right)}>W_\infty,\quad r>r_2.
 \ee
 That is,
 \be\label{4.28}
 W'<N W'<-\frac{W_\infty}r,\quad r>r_2,
 \ee
 since $N<1$ for $r>r_0$. Integrating (\ref{4.28}) leads to
 \be
\nonumber
 W(r;a)<W(r_2;a)-W_\infty\ln\frac{r}{r_2},\quad r>r_2,
 \ee
 so that $W$ fails to stay positive when $r$ is large enough. This contradiction shows that $W_\infty$ can only be zero.
 \medskip

 In summary, we have established that the initial value problem consisting of (\ref{4.1}) and (\ref{4.4}) has a solution
 $W$ satisfying the boundary condition (\ref{4.3}) when (\ref{4.8}) is fulfilled and the initial slope parameter $a$ is suitably chosen in the range
 \be
 a=\frac{2(1-2Q^2\tau)}{(1-Q^2)}<0.
\ee
 Moreover, $W$ enjoys the properties stated in (\ref{4.22}).

%%%%%%%%%%%%%%%%%%%%%%%%%%%%%%%%%%%%%%%%%%%%%%%%%%%%%%%%%%%%%%%%%%%%%%%%%%%%%%%
 \subsection{The numerical results}
%%%%%%%%%%%%%%%%%%%%%%%%%%%%%%%%%%%%%%%%%%%%%%%%%%%%%%%%%%%%%%%%%%%%%%%%%%%%%%%

The solutions of (\ref{4.1}) interpolating between the asymptotics
(\ref{n0}) and (\ref{n1})
are again constructed numerically.
In our numerical approach, we fix $r_H=1$ and  $\tau=1$ as input parameters 
and look for nodeless solutions.
Then (\ref{4.1})  becomes an eigenvalue problem in terms of $Q$.
The corresponding value of $Q$ is found by
using a (numerical) shooting procedure, being uniquelly fixed under these assumptions.

The profile of the  solution for a typical RN background
is shown in Figure 3.
In Figure 4 we show the existence line of solutions in the $(M,Q)$-parameter space.
This corresponds to the critical RN solutions which bifurcate
into a branch of BHs with nA hair.
The hairy solutions are found
when promoting the nA cloud to non-linear level and taking
into account the backreaction \cite{Radu:2011ip}.

%%%%%%%%%%%%%%%%%%%%%%%%%%%%%%%%%%%%%%%%%%%%%%%%%%%%%%%%%%%
 \setlength{\unitlength}{1cm}
\begin{picture}(8,6)
\put(3,0.0){\epsfig{file=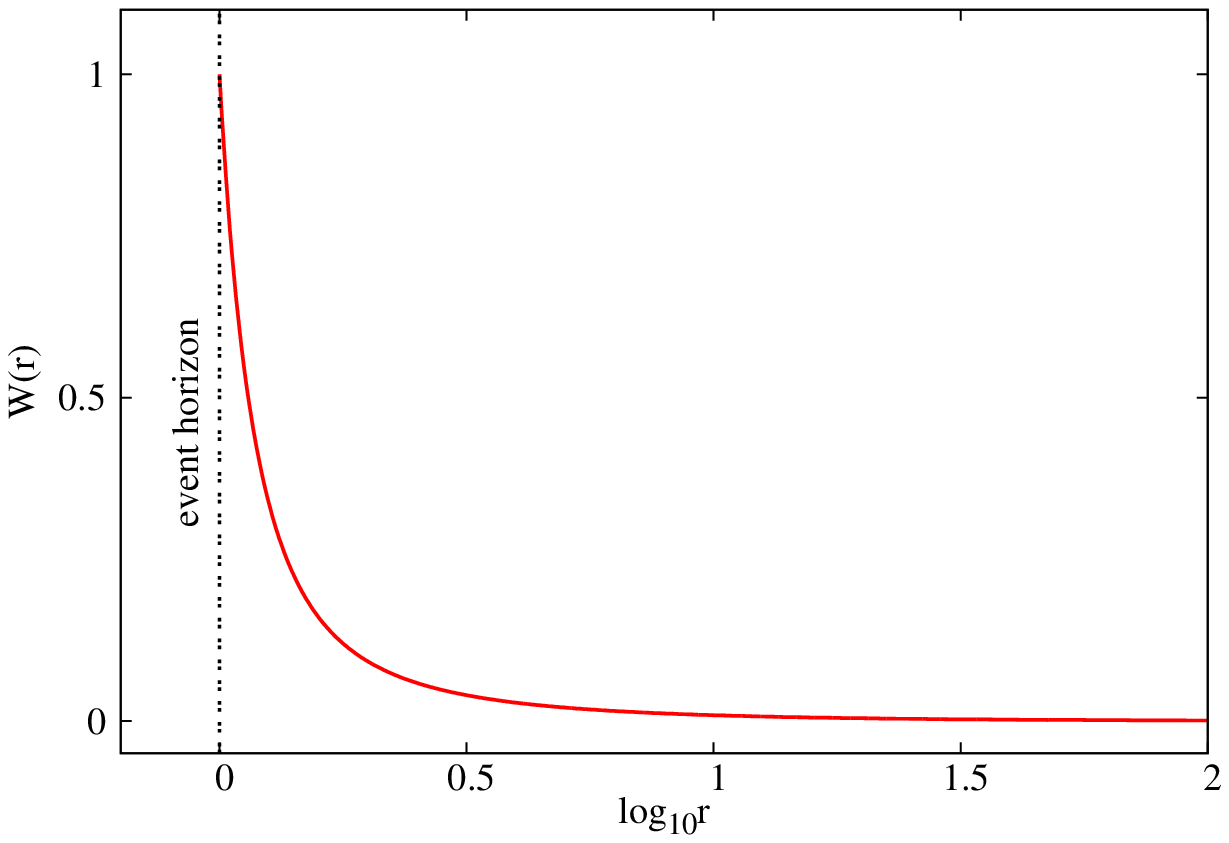,width=8cm}}
\end{picture}
\\
\\
{\small {\bf Figure 3.}
 The profile of a typical solution of the equation (\ref{4.1}).
The parameters of the corresponding RN background are $M=0.8465$, $Q=0.8324$ and $\tau=1.385$.
 }
\vspace{0.5cm}
%%%%%%%%%%%%%%%%%%%%%%%%%%%%%%%%%%%%%%%%%%%%%%%%%%%%%%%%%%%

%%%%%%%%%%%%%%%%%%%%%%%%%%%%%%%%%%%%%%%%%%%%%%%%%%%%%%%%%%%
 \setlength{\unitlength}{1cm}
\begin{picture}(8,6)
\put(3,0.0){\epsfig{file=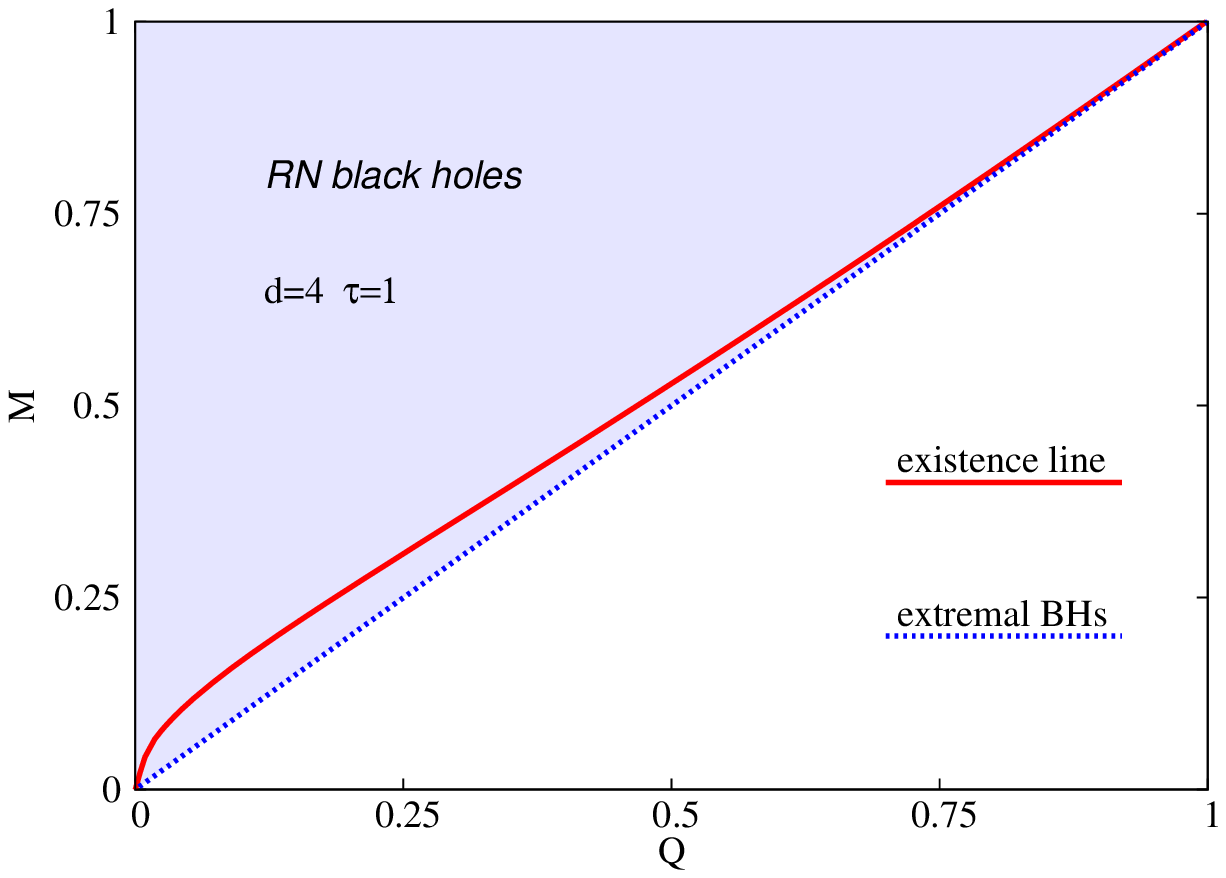,width=8cm}}
\end{picture}
\\
\\
{\small {\bf Figure 4.}
Mass $M$ $vs.$ charge $Q$ for RN black holes in $d=4$ dimensions. The dotted blue
 curve corresponds to extremal BHs
(RN BHs exist above it (shaded region).
The non-Abelian clouds exist along the red line.
 }
\vspace{0.5cm}
%%%%%%%%%%%%%%%%%%%%%%%%%%%%%%%%%%%%%%%%%%%%%%%%%%%%%%%%%%%

%%%%%%%%%%%%%%%%%%%%%%%%%%%%%%%%%%%%%%%%%%%%%%%%%%%%%%%%%%%%%%%%%%%%%%%%%%%%%%%
\section{Higher dimensional generalizations in Minkowskian backgrounds: $d=5$}
%%%%%%%%%%%%%%%%%%%%%%%%%%%%%%%%%%%%%%%%%%%%%%%%%%%%%%%%%%%%%%%%%%%%%%%%%%%%%%%

%%%%%%%%%%%%%%%%%%%%%%%%%%%%%%%%%%%%%%%%%%%%%%%%%%%%%%%%%%%%%%%%%%%%%%%%%%%%%%%
 \subsection{The setting}
%%%%%%%%%%%%%%%%%%%%%%%%%%%%%%%%%%%%%%%%%%%%%%%%%%%%%%%%%%%%%%%%%%%%%%%%%%%%%%%

It is of interest to see if the pattern discussed in the previous two Sections
occurs also in more than four dimensions, $d>4$.
As discussed in \cite{Manvelyan:2008sv},
the higher dimensional (planar) RNAdS black holes
are also unstable when considered as solutions of EYM theory with a negative cosmological constant.
This results as well in the occurrence of a set of critical (embedded Abelian)
BH solutions possessing  nA clouds.

  We expect that this is also the case for the solutions in a Minkowski spacetime
background like that discussed in Section 3.
Indeed, at least in $d=5$ and $d=6$ dimensions,
 this pattern repeats for the Lagrangian (\ref{L}) with gauge group $SO(d+1)$,
the general setup being presented in Appendix A2.
By solving the linearized  YM equation (\ref{4.1-new}),
we have verified that nA clouds around RN black holes exist in those cases.
The overall picture is similar to that discussed in the previous Section
and it will be reported elsewhere~ \cite{to}.
Also, the existence proof given in the previous section can be adapted to the  higher dimensional case.

In what follows, we shall not exercise the option of using higher-order YM terms, but rather will employ 
a Chern-Simons (CS) term in the YM action  (\ref{L}) 
in $lieux$ of the $F(4)^2$ in (\ref{L}) as the simpler,
or less nonlinear alternative~\footnote{The Chern-Simons terms can be employed in all (higher) odd dimensions, but finite energy solutions
exist only for Lagrangians in which the lowest order YM term $F(2p)^2$ is of higher order than the $p=2$ case in (\ref{L}). Of course, the Reissner-Nordstr\"om solutions
in that case would be those of the higher order gravities studied in \cite{Chakrabarti:2001di}. In that case one might also replace the usual Einstein gravity
by its higher order versions employed there.}.
Moreover, this option has a good justification, since
a CS term  appears in various supersymmetric theories,
the ${\cal N}=8,~D=5$  gauged supergravity model
\cite{Gunaydin:1985cu}, \cite{Cvetic:2000nc} being perhaps the best known case,
due to its role in the conjectured AdS/CFT correspondence.

In $d=5$ dimensions, the CS Lagrangian reads
% to the higher order curvature terms of the YM hierarchy,
\begin{eqnarray}
\label{CS5}
 &&{\cal L}_{s} =  \kappa   \epsilon_{I_1\cdots I_6}
\Big(F^{I_1 I_2} \wedge  F^{I_3 I_4}\wedge  A^{I_5 I_6}
  -
 F^{I_1 I_2}\wedge  A^{I_3 I_4}\wedge  A^{I_5 J} \wedge A^{J I_6}
\\
\nonumber
  && {~~~~~~~~~~~~~~~~~~~~~}
 +\frac{2}{5}   A^{I_1 I_2}\wedge  A^{I_3 J}\wedge  A^{J I_4}\wedge
A^{I_5 K} \wedge A^{K I_6}
\Big),
\end{eqnarray}
where $A^{IJ}$ are the $SO(6)$ gauge fields,  $F^{IJ}=dA^{IJ}+ A^{IK}\wedge A^{KJ}$,
and  $\kappa$  the CS coefficient.
%Such  CS  terms appear in various supersymmetric theories and lead to a variety
%of interesting features already in the Abelian case.
%
The gauge fields ansatz contains only two functions, a magnetic gauge potential
$w$, and an $U(1)$ electric one, $V$:
\begin{eqnarray}
\label{YMansatz5}
A=
\frac{w(r )+1}{r} \gamma_{ij}\frac{x^i}{r}dx^j
+V(r )\gamma_{56}dt,~~{\rm with~~} i,j=1, \dots,4\,,
\end{eqnarray}
$\gamma_{ij}$ being the representation matrices of $SO(4)$, and $\gamma_{56}$ of the $SO(2)$, subalgebras
of $SO(6)$.  The Cartesian coordinates $x^i$ are related to the spherical coordinates in (\ref{RN3})
as in flat space.

The CS term (\ref{CS5}) is added to the usual EYM Lagrangian,
the resulting equations being displayed in Appendix A3.
The electrically charged RN BH is a solution of this model, with
\begin{eqnarray}
w(r)\equiv -1, ~~V(r)= \frac{Q}{r^2}-\frac{Q}{r_H^2},
\end{eqnarray}
and a line element
\bea
\label{RN3}
ds^{2}=\frac{dr^{2}}{N(r)}+r^2
d\Omega_3^2
- N(r) dt^{2},~~
{\rm with}~~N(r)=1-\frac{M}{r^2}+\frac{Q^2}{r^4},
\eea
where $M,Q$ are two parameters fixing the mass and electric charge of BH, respectively.
Also, the outer BH horizon is located at $r=r_H=(M+\sqrt{M^2-4Q^2})/2>0$, where $N(r_H)=0$.

Restricting again to an infinitesimally small magnetic field, one takes again
\begin{eqnarray}
w(r)=-1+\epsilon W(r),
\end{eqnarray}
with $W(r)$ a solution of the equation (which results by linearizing the $w$-equation in (\ref{eqs5}))
\begin{eqnarray}
\label{eq2}
r (rNW')'=4(1+\frac{4 Q \kappa}{r^2})W.
\end{eqnarray}

%%%%%%%%%%%%%%%%%%%%%%%%%%%%%%%%%%%%%%%%%%%%%%%%%%%%%%%%%%%%%%
\subsection{The existence of  solutions}
%%%%%%%%%%%%%%%%%%%%%%%%%%%%%%%%%%%%%%%%%%%%%%%%%%%%%%%%%%%%%%

We are interested in smooth solutions
of the above equation
which interpolate between the following
boundary values
\begin{eqnarray}
\label{c1}
W(r_H)=b, ~~ W(\infty)=0.
\end{eqnarray}
\medskip
An approximate solution compatible with (\ref{c1})
can also be constructed;
for $r\to r_H$ one finds
\begin{eqnarray}
\label{51}
W(r)= b+\frac{2b r_H(4\kappa Q+r_H^2)}{r_H^4-Q^2 } (r-r_H)+O(r-r_H)^2
\end{eqnarray}
%where
%\begin{eqnarray}
%w_1=\frac{2b r_H(4\kappa Q+r_H^2)}{r_H^4-Q^2 },
%\end{eqnarray}
with $b\neq 0$ an arbitrary parameter.
Since the eq. (\ref{eq2}) is linear, we set $b=1$
without any loss of generality.
The corresponding expression for large $r$ reads
\begin{eqnarray}
\label{52}
W(r)=\frac{J}{r^2}+\frac{2J(Q^2+2\kappa r_H^2+r_H^4)}{3r_H^2}\frac{1}{r^4}+O(1/r^6),
\end{eqnarray}
%where
%\begin{eqnarray}
%w_2=\frac{2J(Q^2+2\kappa r_H^2+r_H^4)}{3r_H^2},
%\end{eqnarray}
with $J$ a constant.

To simplify notation, we set $r_H=1$. Hence
\be
N(r)=1-\frac{Q^2+1}{r^2}+\frac{Q^2}{r^4}=\frac1{r^4}(r^2-1)(r^2-Q^2),
\ee
and $Q$ satisfies
\be
\nonumber
-1<Q<0,\quad 1+4Q\kappa <0,
\ee
which may be combined into the compressed condition
\be
\nonumber
-1<Q<-\frac1{4\kappa},\quad 0<\kappa<\frac12.
\ee

On the other hand, from (\ref{eq2}) we have
\be\label{3.4}
W'(r)=\frac{4r^3}{(r^2-1)(r^2-Q^2)}\int_1^r\frac1{\rho}\left(1+\frac{4Q\kappa}{\rho^2}\right)W(\rho)\, d\rho,\quad r>1.
\ee
Using the L'Hopital's rule we find
\be
\nonumber
b(Q)\equiv W'(1)=\lim_{r\to1} W'(r)=\frac2{1-Q^2}(1+4Q\kappa)<0.
\ee
We  note that
\be\label{3.6}
\lim_{Q\to-\frac1{4\kappa}}b(Q)=0,\quad\lim_{Q\to-1}b(Q)=-\infty,
\ee
and
\be
b'(Q)=\frac4{(1-Q^2)^2}(2\kappa Q^2+Q+2\kappa),
\ee
which is positive if $\kappa$ satisfies the additional condition $\kappa>\frac14$. Hence, we collect our condition on $\kappa$ here:
\be
\frac14<\kappa<\frac12.
\ee
Under this condition the function $b(Q)$ strictly increases in the interval $(-1,-\frac1{4\kappa})$ with values from $-\infty$ to zero. Such a property is essential for
the construction of our solution to follow.

In (\ref{3.4}) the weight function changes its sign at $\rho=r_0$ where
\be
r_0=2\sqrt{-Q\kappa}>1.
\ee
So we first study the properties of the local solution of (\ref{3.4}) in $I_0=[1,r_0]$.

We note that $W(r)>0$ when $r>1$ and $r$ is close to $1$. Thus, as far as $W$ stays non-negative and $r\in I_0$, we have $W'(r)<0$. Thus, in this interval, we
have $0\leq W(r)<1$ and so
\bea\label{3.10}
W'(r)&>&\frac{4r^3}{(r^2-1)(r^2-Q^2)}\int_1^r\frac1{\rho}\left(1+\frac{4Q\kappa}{\rho^2}\right)\, d\rho\nn\\
&=&\frac{4r^3}{(r^2-1)(r^2-Q^2)}\left(\ln r+\frac{r_0^2}2\left[\frac1{r^2}-1\right]\right)\nn\\
&>&\frac{2r^3 r_0^2}{(r^2-1)(r^2-Q^2)}\left(\frac1{r^2}-1\right)\nn\\
&>&-\frac{2r_0^2 }{r(1-Q^2)},\quad r\in I_0.
\eea
Integrating (\ref{3.10}) and applying the condition $W(1)=1$, we obtain
\bea\label{3.11}
W(r)&>&1-\frac{2r_0}{1-Q^2}\ln r_0\nn\\
&=&1-\frac{2\sqrt{-Q\kappa}}{1-Q^2}\ln\left(-4 Q\kappa\right),\quad r\in I_0.
\eea
It is clear that when $Q$ is close to $-\frac1{4\kappa}$ the right-hand side of (\ref{3.11}) stays positive. This proves that the set
\be
\nonumber
B^+=\left\{Q\in\left(-1,-\frac1{4\kappa}\right)\,\bigg|\,W(r)>0\mbox{ for all }r\in I_0\right\}
\ee
where $W$ solves the initial value problem
\be\label{3.13}
 W\mbox{ satisfies } (\ref{eq2})~\mbox{ and the initial condition }~ W(1)=1,~W'(1)=b(Q)
\ee
is nonempty. The continuous dependence of the solution to the initial condition and the parameters in the differential equation then indicates that $B^+$ is open.

Similarly, it may be shown that the set
\be
\nonumber
B^-=\left\{Q\in\left(-1,-\frac1{4\kappa}\right)\,\bigg|\,W(r)<0\mbox{ for somel }r\in I_0\right\}
\ee
is nonempty, where $W$ solves the initial value problem (\ref{3.13}). To see this we may choose $Q$ to be close to $-1$ which makes $W$ to start with
a sufficiently negtive slope in view of (\ref{3.6}) and drives $W$ to the negative value range quickly. Of course $B^-$ is also open.

By the connectedness of the interval $\left(-1,-\frac1{4\kappa}\right)$ we see that
\be
\nonumber
B^0=\left(-1,-\frac1{4\kappa}\right)\setminus(B^+\cup B^-)\neq\emptyset.
\ee

For $Q\in B^0$, we see that $W(r)\geq0$ for all $r\in I_0$. Since $W'(r)<0$ for $r\in I_0$, we must have $W(r_0)=0$.

In summary, we have obtained some $Q\in\left(-1,-\frac1{4\kappa}\right)$ so that the solution of (\ref{3.13}) satisfies
\be\label{3.16}
W(r)>0\quad\mbox{for } r_0\in[1,r_0); \quad W(r_0)=0;\quad W'(r)<0,\,\, r\in[1,r_0].
\ee

Let $W$ be a solution of (\ref{3.13}) satisfying (\ref{3.16}). Then $W(r)<0$ for $r>r_0$ but $r$ is close to $r_0$ since $W(r_0)=0$ and $W'(r_0)<0$.
For $r>r_0$ we rewrite (\ref{3.4}) as
\be
\nonumber
W'(r)=\frac{4r^3}{(r^2-1)(r^2-Q^2)}\left(\int_1^{r_0}+\int_{r_0}^r\right)\frac1{\rho}\left(1-\frac{r_0^2}{\rho^2}\right)W(\rho)\, d\rho,\quad r>r_0.
\ee
We see that $W'(r)$ for all $r>r_0$, in the interval of existence, since $W(r)<0$ beyond $r_0$. In particular, we have established that the set
\be
\nonumber
{\cal Q}^-=\left\{Q\in\left(-1,0\right]\,\bigg|\, W'(r)<0\mbox{ for all }r>1\mbox{ and }W(r)<0\mbox{ for some }r>1\right\}
\ee
is not empty. This set is readily seen to be open.

To proceed further, we define
\be
\nonumber
{\cal Q}^+=\left\{Q\in\left(-1,0\right]\,\bigg|\, W'(r)>0\mbox{ for some }r>1\right\}.
\ee
Then ${\cal Q}^+\neq\emptyset$. In fact, we may formally rewrite the local solution of (\ref{eq2}) in the form (\ref{3.4}) and use the L'Hopital's rule to find
$W'(1)$ again as before. In particular, for $Q<0$ and close to $Q=0$, we have $W'(1)>0$. Thus $W'(r)>0$ for $r$ near $r=1$ and $r>1$. It is obvious that
${\cal Q}^+$ is also open.

Now the connectedness of $(-1,0]$ shows that
\be
\nonumber
{\cal Q}^0=(-1,0]\setminus({\cal Q}^+\cup{\cal Q}^-)\neq\emptyset.
\ee
Take $Q\in {\cal Q}^0$. Then $Q<0$. For such $Q$, let $W$ be the corresponding solution of (\ref{3.13}) where we no longer require $Q$ to be confined within
the interval $\left(-1,-\frac1{4\kappa}\right)$ but within $(-1,0)$. For such $W$, we have the properties
\be\label{3.21}
W'(r)\leq0,\quad W(r)\geq0,\quad r>1.
\ee
It is clear that in fact $W(r)>0$ for all $r>1$. Indeed, if there is some $r_1>1$ such that $W(r_1)=0$, then $W$ attains it interior minimum at $r_1$ such that
$W'(r_1)=0$. Applying the uniqueness theorem of the initial value problem of ordinary differential equations we derive $W\equiv0$ which is false.

We claim $W'(r)<0$ for all $r>1$. In fact since $W(r)>0$ we have $W'(r)<0$ for $r\in I_0$ in view of (\ref{3.4}). Assume otherwise that $W'(r_1)=0$ at some
$r_1>r_0$ and that $r_1$ is the left-most such point. Then (\ref{3.4}) gives us
\be
\nonumber
\int_1^{r_1}\frac1{\rho}\left(1+\frac{4Q\kappa}{\rho^2}\right)W(\rho)\, d\rho=0.
\ee
Hence we have
\bea
W'(r)&=&\frac{4r^3}{(r^2-1)(r^2-Q^2)}\left(\int_1^{r_1}+\int_{r_1}^r\right)\frac1{\rho}\left(1-\frac{r_0^2}{\rho^2}\right)W(\rho)\, d\rho\nn\\
&=&\frac{4r^3}{(r^2-1)(r^2-Q^2)}\int_{r_1}^r\frac1{\rho}\left(1-\frac{r_0^2}{\rho^2}\right)W(\rho)\, d\rho>0,\quad r>r_1,
\eea
since $W(r)>0$. This is false. Therefore we arrive at the following slightly strengthened version of (\ref{3.21}):
\be\label{3.24}
W'(r)<0,\quad W(r)>0,\quad r>1.
\ee

With the properties of $W$ established, we see in particular that
\be
W_\infty\equiv \lim_{r\to\infty}W(r)
\ee
exists and $W_\infty[0,1)$. We further claim $W_\infty=0$. Otherwise, if $W_\infty>0$, then we have
\be
\nonumber
0<W_\infty<W(r)<1,\quad r>1.
\ee
Inserting this into (\ref{3.4}) we see that there are constants $C_0,C_1,C_2>0$ such that
\be
\nonumber
W'(r)>\frac{C_0}r\left({C_1}\ln r-C_2\right),\quad r>r_2
\ee
where $r_2>1$ is sufficiently large. In particular, $W'(r)>0$ when $r$ is large enough which is false again.

In summary, we have shown the existence of a solution $W$ of (\ref{eq2}) which enjoys the properties
\be
W(1)=1;\quad W'(r)<0,\quad W(r)>0,\quad 1<r<\infty;\quad W(\infty)=0,
\ee
provided that $Q\in(-1,0)$ is suitably chosen and $\kappa$ stays in the interval $(\frac14,\frac12)$.

%%%%%%%%%%%%%%%%%%%%%%%%%%%%%%%%%%%%%%%%%%%%%%%%%%%%%%%%%%%%%%%%%%%%%%%%%%%%%%%
 \subsection{The numerical results}
%%%%%%%%%%%%%%%%%%%%%%%%%%%%%%%%%%%%%%%%%%%%%%%%%%%%%%%%%%%%%%%%%%%%%%%%%%%%%%%

The smooth solution interpolating between (\ref{51}) and (\ref{52})
is constructed again by numerical integration.
Not completely unexpected,
the picture here is similar to that found for the model in Section 3.
In Figure 5 we show the existence line found for $\kappa=0.6$.
The profile of a typical solution of (\ref{eq2})
is similar to that in Figure 3 and we shall not display here.

%%%%%%%%%%%%%%%%%%%%%%%%%%%%%%%%%%%%%%%%%%%%%%%%%%%%%%%%%%%
 \setlength{\unitlength}{1cm}
\begin{picture}(8,6)
\put(3,0.0){\epsfig{file=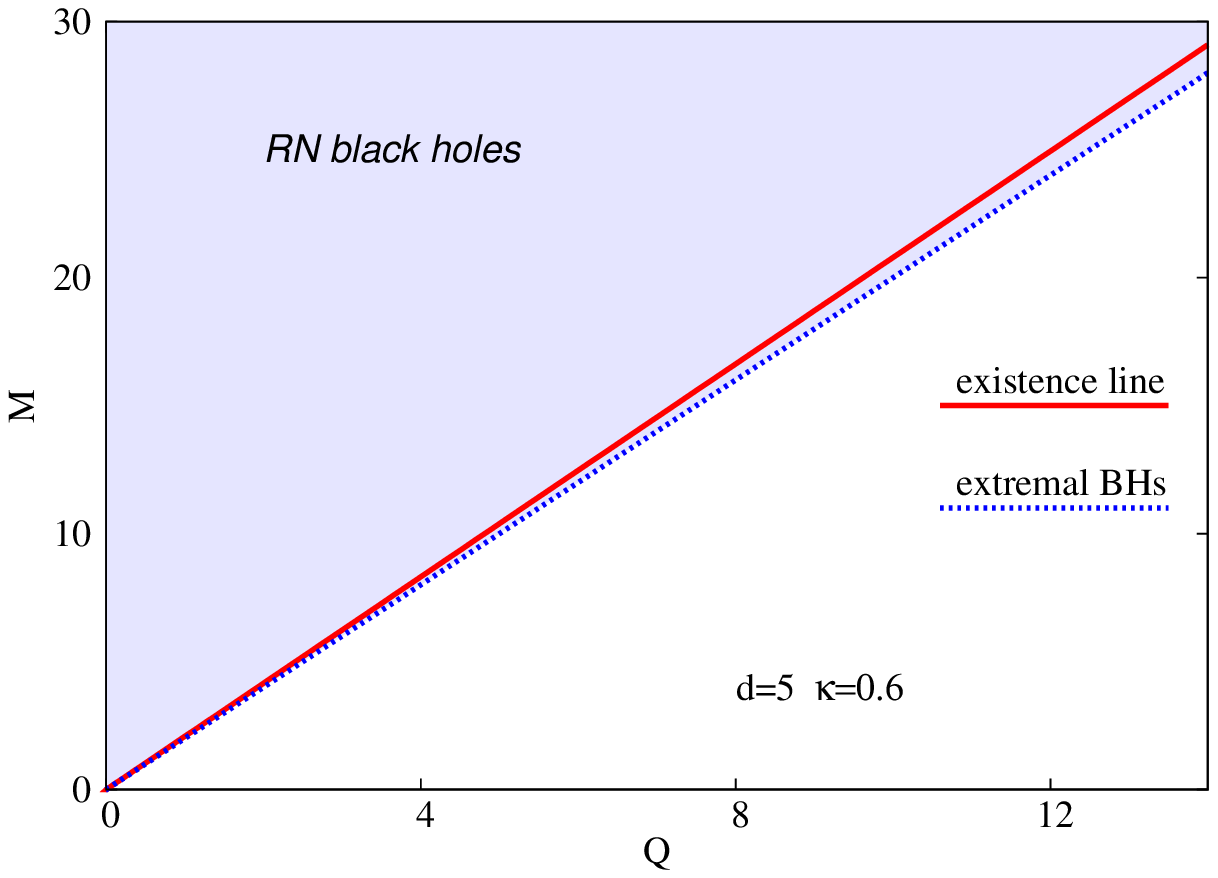,width=8cm}}
\end{picture}
\\
\\
{\small {\bf Figure 5.}
Mass $M$ $vs.$ charge $Q$ for RN black holes in $d=5$
dimensions. The dotted blue
 curve corresponds to extremal BHs
(RN BHs exist above it (shaded region)).
The non-Abelian  clouds exists along the red line.
 }
\vspace{0.5cm}
%%%%%%%%%%%%%%%%%%%%%%%%%%%%%%%%%%%%%%%%%%%%%%%%%%%%%%%%%%%

As discussed in
\cite{Brihaye:2010wp},
\cite{Brihaye:2011nr},
these nA clouds may be continued past the
point where they are an infinitesimally small perturbation of RN BHs.
This results in a set of black holes
with magnetic nA hair,
which possess also a finite mass and electric charge.
Remarkably,  some of these nA configurations are stable under linear,
spherically symmetric perturbations.

%%%%%%%%%%%%%%%%%%%%%%%%%%%%%%%%%%%%%%%%%%%%%%%%%%%%%%%%%%%%%%%%%%%%%%%%%%%%%%%
\section{Conclusions}
%%%%%%%%%%%%%%%%%%%%%%%%%%%%%%%%%%%%%%%%%%%%%%%%%%%%%%%%%%%%%%%%%%%%%%%%%%%%%%%

One of the most interesting developments in the physics of AdS spacetime was
the discovery in 2008 by Gubser of the $p$-wave
 {\it ``holographic superconductors''}.
The starting point there was the observation that RNAdS
BH becomes unstable
 when considered as a solution of the Einstein--Yang-Mills theory,
with the appearance of a magnetic nA cloud close
to the horizon.
This feature occurs for a
particular set of RNAdS
configurations which form a line in the parameter space,
the hairy BHs being
the nonlinear realization of those {\it marginally stable solutions}. 

%\medskip
The main purpose of this work was to show that, despite the
different asymptotic structure of spacetime and the different
horizon topology, the asymptotically flat RN black holes can also become unstable
in the presence of nA gauge fields,
with the occurrence of nA clouds around the horizon.
 We provided two examples of this, for $d=4$ and $d=5$ spacetime dimensions, 
in Sections {\bf 3} and {\bf 4} respectively.

In contrast to the case of AdS background, 
where one finds EYM solutions supporting the electric component of the nA $SU(2)$ connection $A_t$ in $d=4$
spacetime, 
the existence of such nA hairy solutions in the case of flat background requires two crucial ingredients.  
The first one is that the  gauge group should be large enough
(at least $SO(d+1)$ for our models, with $d$ the spacetime dimension).
The second ingredient is  the presence of some nonlinear terms in the nA connection and curvature, of higher order than
the usual (quadratic) $F^2$ one. The higher order terms used here are variously, the (quartic) $F^4$ and the Chern-Simons terms.
In both these two (flat background) cases, analytic proofs for the existence of the perturbed solutions were given in addition to
numerical construction.% as in the AdS example in Section~{\bf 2}. 

The general mechanism appears to be the following: in all cases,
the RN BH remains a(n embedded Abelian) solution of the full model.
However, for some range of the parameters, the extra terms mentioned above give
a tachyonic mass for the vacuum perturbations
of the nA magnetic fields around the Abelian solutions,
with the appearance of a nA condensate.
Similar to the AdS case, this implies
the  occurrence  of a branch of fully nA BHs,
which are generically thermodynamically favored over the Abelian
configurations.

It would be interesting to further pursue the unveiled similarity of  the Minkowskian case here 
with the AdS case, and to
investigate the possible relevance of these aspects in providing
analogies to phenomena observed in condensed matter physics.

%%%%%%%%%%%%%%%%%%%%%%%%%%%%%%%%%%%%%%%%%%%%%%%%%%%%%%%%%%%%%%%%%%%%%%%%%%%%%%
\section*{Acknowledgements}
%%%%%%%%%%%%%%%%%%%%%%%%%%%%%%%%%%%%%%%%%%%%%%%%%%%%%%%%%%%%%%%%%%%%%%%%%%%%%%
 E.R.~gratefully acknowledges funding from the FCT-IF programme. 
This work was partially supported by the H2020-MSCA-RISE-2015 Grant No. StronGrHEP-690904, 
and by the CIDMA project UID/MAT/04106/2013.

	\medskip
	\medskip
	
%%%%%%%%%%%%%%%%%%%%%%%%%%%%%%%%%%%%%%%%%%%%%%%%%%%%%%%%%%%%%%%%%%%%%%%%%%%%%%%%%%
%\newpage
\appendix

%%%%%%%%%%%%%%%%%%%%%%%%%%%%%%%%%%%%%%%%%%%%%%%%%%%%%%%%%%%%%%%%%%%%%%%%%
\section{The nonlinear problems}
%%%%%%%%%%%%%%%%%%%%%%%%%%%%%%%%%%%%%%%%%%%%%%%%%%%%%%%%%%%%%%%%%%%%%%%%%

%%%%%%%%%%%%%%%%%%%%%%%%%%%%%%%%%%%%%%%%%%%%%%%%%%%%%%%%%%%%%%%%%%%%%%%%%
\subsection{The EYM-AdS system in $d=4$ dimensions}
\setcounter{equation}{0}
\renewcommand{\theequation}{A.\arabic{equation}}
%%%%%%%%%%%%%%%%%%%%%%%%%%%%%%%%%%%%%%%%%%%%%%%%%%%%%%%%%%%%%%%%%%%%%%%%%	
We consider a general class of EYM-AdS solutions  
with a line element  
\begin{eqnarray}
\label{metric-AdS}
ds^{2}=\frac{dr^{2}}{N(r)}+r^2 (d\theta^2+f_k^2(\theta) d \varphi^2)- N(r)\sigma^2(r) dt^{2},~~
{\rm with}~~N(r)=k-\frac{2m(r)}{r}+\frac{r^2}{L^2},
\end{eqnarray}
where $k=0,\pm 1$ and
\begin{equation}
f_k(\theta)=\left \{
\begin{array}{ll}
\sin\theta, & {\rm for}\ \ k=1 \\
\theta , & {\rm for}\ \ k=0 \\
\sinh \theta, & {\rm for}\ \ k=-1.
\end{array} \right.
\end{equation}
For any value of $k$, the metric (\ref{metric-AdS}) possesses the same amount of symmetries, since 
  $ d\theta^{2}+f_k^{2}(\theta) d\varphi^{2}$
is the line element on a two-dimensional surface of constant curvature $2k$.
The spherically symmetric solutions have $k=1$, in which case one can find both solitons and black hole solutions \cite{Winstanley:1998sn};
for $k=0,-1$ one finds only ``topological'' black holes 
(see \cite{Baxter:2015tda} 
for recent results on such configurations).

The $SU(2)$-YM Ansatz compatible with the symmetries of the line-element (\ref{metric-AdS})
has been proposed in  
\cite{VanderBij:2001ia}, 
\cite{Mann:2006jc},
and reads
\begin{equation} 
\label{A}
A=\frac{1}{2 } \bigg[ V(r ) \tau_3 dt+ 
  w(r ) \tau_1  d \theta
+\left(\frac{1}{f_k(\theta)}\frac{f_k(\theta)}{d \theta} \tau_3
+ f_k(\theta) w(r ) \tau_2  \right)  d \varphi 
\bigg].
\end{equation}

The gauge potentials $w(r)$, $V(r)$
and the metric functions $m(r)$, $\sigma(r)$
solve the EYM equations
\begin{eqnarray}
\label{eqs-AdS}
&&
m'=
Nw'^2
+\frac{r^2V'^2}{2\sigma^2}
+\frac{(w^2-k)^2}{2r^2}
+\frac{w^2V^2}{N\sigma^2},
~~
\sigma'=\frac{2\sigma}{r}(w'^2+\frac{w^2V^2}{N^2\sigma^2}),
\\
&&
\nonumber
w''
+(\frac{N'}{N}+\frac{\sigma'}{\sigma})w'
+\frac{w(k-w^2)}{r^2N}+\frac{w V^2}{N^2\sigma^2}=0,
~~
V''+(\frac{2}{r}-\frac{\sigma'}{\sigma})V'-\frac{2V}{r^2N}w^2=0.
\end{eqnarray} 
One can easily see that
  $k=0$ is special,
since the RN black hole  with a nonzero electric field and a vanishing magnetic flux 
is a solution  in that case only. 
For $k=1$, the magnetic field trivializes for $w(r)=\pm 1$.
However, from the $w$-equation, this is not consistent with
keeping a nonzero electric potential $V(r)$.
Moreover,
the solutions with $k=-1$
do not possess a vacuum,
since one cannot find a real value of $w(r)$ which gives a vanishing magnetic field
(even for $V(r)\equiv 0$).

%%%%%%%%%%%%%%%%%%%%%%%%%%%%%%%%%%%%%%%%%%%%%%%%%%%%%%%%%%%%%%%%%%%%%%%%%
\subsection{The EYM-$F^4$ system  in $d$-spacetime dimensions}
%\setcounter{equation}{0}
%\renewcommand{\theequation}{B.\arabic{equation}}
%%%%%%%%%%%%%%%%%%%%%%%%%%%%%%%%%%%%%%%%%%%%%%%%%%%%%%%%%%%%%%%%%%%%%%%%%	
We consider the following action in $d$-dimensions  (with $\tau_1$, $\tau_2$ positive constants):
\begin{eqnarray}
\label{action-p2}
S=\int d^d x
\sqrt{-g}
\bigg [
\frac{1}{4}R
-\frac{1}{2}\tau_1 \mbox{Tr}  \{F(2)^2 \}
+\frac{3}{2}\tau_2
 \mbox{Tr}  \{F(4)^2 \} \bigg ],
\end{eqnarray}	
describing Einstein gravity coupled with 
the first two terms in the Yang-Mills hierarchy~\footnote{$F(2p)^2$ having been proposed as systems supporting instantons on
$\R^{4p}$ in~\cite{Tchrakian:1984gq}}, considered in \cite{Brihaye:2002hr}.
In the above relation,	
$F(2)=F_{\mu\nu}$ denotes the Yang-Mills curvature $2$-form, while
$F(4)=F_{\mu\nu\rho\si}$ is the Yang-Mills $4$-form resulting from the total antisymmetrisation of $F(2)$.
The $4$-form  $F_{\mu\nu\rho\si}$  can be expressed conveniently as
 \bea
F_{\mu\nu\rho\si}=\{F_{\mu[\nu},F_{\rho\si]}\},~
\label{p=2} 
\eea
where $\{F_{\mu\nu},F_{\rho\si}\}$ denotes the anticommutator of $F_{\mu\nu}$ and $F_{\rho\si}$, while the notation
$F_{\mu[\nu}\,F_{\rho\si]}$ implies the cyclic symmetry on $(\nu,\rho,\si)$, $i.e$
%\[
$
F_{\mu\nu}\,F_{\rho\si}+F_{\mu\rho}\,F_{\si\nu}+F_{\mu\si}\,F_{\nu\rho}\,.
$
%\]
This results in 
\be
\label{F^4}
\mbox{Tr}  \{F(4)^2 \} =6\,\mbox{Tr}\left[(F_{\mu\nu}F_{\rho\si})^2-4(F_{\mu\rho}F_{\rho\nu})^2+(F_{\mu\nu}^2)^2\right],
\ee
which is analogous to the corresponding expression for the Gauss-Bonnet term in gravity.
 Unlike the latter however, \re{F^4} is not a total divergence and hence 
does not trivialize in $d=4$ dimensions,   
as long as the component $A_0$ of the YM connection is supported, which is achieved by choosing
a large enough gauge group. 
In that case it can be expressed in the equivalent form displayed in eq. (\ref{L}).

A convenient parametrization of a spherically symmetric line element is %in $d-$dimensions is
	\begin{eqnarray}
\label{metric-d}
ds^{2}=\frac{dr^{2}}{N(r)}+r^2 d\Omega_{d-2}^2- N(r)\sigma^2(r) dt^{2},~~
{\rm with}~~N(r)=1-\frac{2m(r)}{r^{d-3}},
\end{eqnarray}
%in terms of two functions, $m(r)$ and $\sigma(r)$,
 with $m(r)$  the mass function and $d\Omega_{d-2}^2$ the metric on the $(d-2)$-sphere. 

In $d-$spacetime dimensions,
the  minimal gauge group allowing a spherically symmetric 
nA Ansatz containing both electric and magnetic parts is $SO(d+1)$.
In what follows, we shall restrict to a consistent  $SO(d-1)\times SO(2)$ truncation of the general Ansatz,
parametrised by two potentials, a magnetic one $w(r)$ and an electric one
 $V(r)$:
\begin{eqnarray}
\label{YMansatz-d}
A= 
\frac{w(r)+1}{r}\ \gamma_{ij}\frac{x^i}{r}dx^j
+V(r)\gamma_{d,d+1}dt,~~{\rm with~~} i,j=1, \dots,d-1\,,
\end{eqnarray}
$\gamma_{ij}$ being the representation matrices of $SO(d-1)$, and $\gamma_{d,d+1}$ of the $SO(2)$, subalgebras in $SO(d+1)$.
The Cartesian coordinates $x^i$ are related to the spherical coordinates in (\ref{metric-d})
as in flat space. The matrices $\gamma_{i}$ used here
symbolise the $(d-1)-$dimensional Dirac gamma matrices $\Ga_i$ when $d$ is
odd, and the chiral matrices $(\Si_i,\tilde\Si_i)$  when $d$ is even.

A straightforward computation leads to the following equations of the model\footnote{Note that,
in order to simplify the relations, a factor of $d$
has been absorbed in the expression of $\tau_1$, $\tau_2$.}
%
%%%%%%%%%%%%%%%%%%%%%%%%%%%%%%%%%%%%%%%%%
\begin{eqnarray}
\nonumber
&&m'= r^{d-2}
\bigg\{
\tau_1 \bigg[
\frac{(d-2)}{2}
           \left(
					\frac{Nw'^2}{r^2}+\frac{1}{2}(d-3) W
           \right)
					+\frac{V'^2}{2\sigma^2}
					\bigg ]
					+\tau_2 W 
					        [
									\frac{V'^2}{\sigma^2}+3(d-4) (\frac{Nw'^2}{r^2}+\frac{1}{4}(d-5) W)
									]
\bigg \},
\\
\nonumber
&&
\sigma'=\frac{ \sigma}{r}w'^2\bigg(
                            \tau_1( d-2)+6\tau_2 (d-4) W
												 \bigg),
\\
&&
w''\left(\tau_1+\frac{6(d-4)}{(d-2)} \tau_2 W \right)
+
\tau_1
\bigg [
\left(
\frac{N'}{N}+\frac{\sigma'}{\sigma}+\frac{d-4}{r}
\right)w'			
+\frac{(d-3)}{r^2N}w(1-w^2)
\bigg ]
\\
\nonumber
&&
{~~~ }
+\frac{\tau_2}{(d-2) }
\bigg[
\frac{6(d-4)}{r^4}
         \bigg(
				8 w(w^2-1)w'^2+ (\frac{N'}{N}+\frac{\sigma'}{\sigma}+\frac{d-8}{r})r^4W w'+
\frac{(d-5)}{2r^2N}(1-w^2)^3 
          \bigg)
\\
\nonumber
&&
{~~~ }
+w(1-w^2)\frac{1}{r^2}
                \left(
			3(d-4)((d-5)W+\frac{4Nw'^2}{r^2}) -\frac{4V'^2}{\sigma^2} 
								\right)
\bigg]=0,
\\
\nonumber
&&
\left(V' \frac{r^{d-2}}{\sigma} \big(\tau_1+\frac{2\tau_2(1-w^2)^2}{r^4}\big)\right)'=0.			
 \end{eqnarray}
The equations for the $SU(2)\times U(1)$ ($i.e.$  $SO(3)\times SO(2)$) model introduced in Section 3 are found by taking 
$\tau_1=1$,
$\tau_2=\tau$,
 $d=4$
in the above relations.
Also, one can easily verify that the embedded Abelian RN BH in $d$-dimensions  as given by
\begin{eqnarray}
m(r)=M-\frac{(d-3)Q^2\tau_1}{3r^{d-3}},~~
\sigma(r)=1,~~w(r)=\pm1,~~V(r)=\frac{Q}{r^{d-3}}+c_0, 
\end{eqnarray}
(with $M$, $Q$ two constants fixing the mass and electric charge) 
is a solution.

Again, an infinitesimally small magnetic field is turned on by taking $w(r)=-1+\epsilon W(r)$. 
To leading order, the equation satisfied by $W(r)$ is
\begin{eqnarray}
\label{4.1-new}
(r^{d-4}NW')'=\frac{2(d-3)W}{r^{6-d}}\left(\tau_1-\frac{4(d-3)\tau_2}{(d-2)r^2(d-3)} \right).
\end{eqnarray}

%%%%%%%%%%%%%%%%%%%%%%%%%%%%%%%%%%%%%%%%%%%%%%%%%%%%%%%%%%%%%%%%%%%%%%%%%
\subsection{The EYM-CS system in $d=5$ dimensions}
%\setcounter{equation}{0}
%\renewcommand{\theequation}{C.\arabic{equation}}
%%%%%%%%%%%%%%%%%%%%%%%%%%%%%%%%%%%%%%%%%%%%%%%%%%%%%%%%%%%%%%%%%%%%%%%%%	
We consider the following action
\begin{eqnarray}
\label{action}
S=\int d^5 x
\sqrt{-g}
\bigg [
\frac{1}{4}R-\frac{1}{2 }{\rm Tr}\left \{ F_{\mu\nu}F^{\mu\nu} \right\}
+\kappa_{}
\vep^{\la\mu\nu\rho\si}\mbox{Tr}\,
\big\{A_{\la}\big (F_{\mu\nu}F_{\rho\si}- F_{\mu\nu}A_{\rho}A_{\si}+
\frac25 _{~} A_{\mu}A_{\nu}A_{\rho}A_{\si}\big)
\big\}
\bigg].
\end{eqnarray}	
	The expression of a $d=5$ spherically symmetric line element reads
	\begin{eqnarray}
\label{metric-d4}
ds^{2}=\frac{dr^{2}}{N(r)}+r^2 d\Omega_3^2- N(r)\sigma^2(r) dt^{2},~~
{\rm with}~~N(r)=1-\frac{m(r)}{r^2}.
\end{eqnarray}
For the gauge group $SO(6)$, the corresponding spherically symmetric, time independent
YM Ansatz is given by (\ref{YMansatz5}), in terms of a magnetic
	potential $w(r)$
	and an electric one $V(r)$.
Then the equations of the model are
\begin{eqnarray}
\label{eqs5}
&&m'=\frac{1}{2}
\left(
3r \bigg(
N w'^2
+\frac{(w^2 -1)^2}{r^2}
\bigg)
+\frac{r^3}{\sigma^2}  V'^2
\right),
~~\frac{\sigma'}{\sigma}=\frac{3 w'^2 }{2r},
\\
\nonumber
&&(r\sigma Nw')'=
\frac{2 \sigma  w(w^2 -1)}{r }
+8\kappa
 V'(w^2 -1) ,
~~~\big (\frac{r^3 V'}{\sigma}\big )'= 24 \kappa (w^2 -1)w'.
\end{eqnarray}

%%%%%%%%%%%%%%%%%%%%%%%%%%%%%%%%%%%%%%%%%%%%%%%%%%%%%%%%%%%%%%%%%%%%%%%%%	

\end{document}